\documentclass[runningheads]{llncs}
\usepackage[T1]{fontenc}
\usepackage{graphicx}
\usepackage{amsmath}
\begin{document}
\title{Multilingual Speech Recognition Using Discrete Tokens with a Two-step Training Strategy}
%
\titlerunning{Multilingual Speech Recognition Using Discrete Tokens}
%
\author{Zehan Li\inst{1} \and
Yan Yang\inst{1} \and
Xueqing Li\inst{1,2} \and
Jian Kang\inst{1} \and
Xiao-Lei Zhang\inst{1,2} \and
Jie Li\inst{1}}
\authorrunning{Z. Li et al.}
\institute{Institute of Artificial Intelligence (TeleAI), China Telecom, China \and
Northwestern Polytechnical University, Xi'an, China\\
\email{\{lizh85, yangy142, kangj30, lij86\}@chinatelecom.cn, lixueqing@mail.nwpu.edu.cn, xiaolei.zhang@nwpu.edu.cn}}

\maketitle              
\begin{abstract}
Pre-trained models, especially self-supervised learning (SSL) models, have demonstrated impressive results in automatic speech recognition (ASR) task. While most applications of SSL models focus on leveraging continuous representations as features for training downstream tasks, the utilization of discrete units has gained increasing attention in recent years owing to its lower storage requirements and broader range of applications. In multilingual ASR tasks, representations at different layers of the model contribute differently to various languages, complicating the unification of discrete unit modeling. In this paper, we propose a two-stage training strategy to improve the discrete token performance of pre-trained models and narrow the gap with continuous representation performance. We validate our method on the XLS-R model following the settings of Interspeech2024 Speech Processing Using Discrete Speech Unit Challenge. Our method demonstrates a significant improvement on the ML-SUPERB dataset, achieving a 44\% relative reduction on CER for the XLS-R model. This surpasses the previous baseline set by the WavLM model, which achieves a 26\% relative reduction on CER. Furthermore, our method achieves the first place among all the single-system results on the leaderboard.

\end{abstract}
\section{Introduction\vspace{-0.4cm}}

Conventional end-to-end ASR has made significant improvement over the past decade \cite{watanabe2017hybrid,gulati20_interspeech,gao22b_interspeech,yao2024zipformer}. These models typically use Mel Frequency
Cepstral Coefficients (MFCC) or log Mel filter banks (FBANK) as feature input and are trained with labeled data. In recent years, self-supervised models \cite{baevski2020wav2vec,hsu2021hubert,baevski2022data2vec} trained with a large amount of unlabeled data have achieved impressive results in various downstream tasks. Representations from SSL models also stand out for their versatility and perform better than traditional features on speech-related downstream tasks. However, these representations usually have large dimensions, which takes up a lot of storage space and slows down training process.

Recently, some studies have explored the application of discrete units in downstream tasks such as ASR, text-to-speech and speech translation \cite{yang2023towards,lee2021textless}. This method substantially reduces data storage while maintaining predictive performance comparable to traditional features and allows researchers to model both speech and semantic units, which can bring speech abilities to Large Language Models (LLM) \cite{zhang2023speechgpt,wang2023viola}. 

While most studies focus on optimizing the utilization of discrete tokens, some research has investigated methods to obtain higher-quality discrete tokens for specific downstream tasks. De-duplication and byte pair encode (BPE) strategies are added after extracting discrete tokens \cite{wu2023wav2seq}, which reduces the bit rate of the model and improves the performance of discrete tokens on downstream tasks. 
For multilingual ASR tasks, continuous representations of various SSL models trained with different objective can achieve good performance. However, the contribution of different layers of representation in the pre-trained model is different, resulting in a huge performance gap between discrete tokens and continuous representations for different languages or tasks.

In this paper, we propose a two-stage training strategy to utilize the representations from all layers for acquiring discrete tokens. In the first stage, we use the continuous representations for downstream task training to determine the contribution weight of each layer. In the second stage, we fix these weights to extract discrete tokens, which are then used for training downstream discrete ASR tasks. In order to break the auto-encoder style behavior in the final few layers \cite{pasad2021layer} and make the representation better at encoding word identity, we also apply a fine-tuning strategy with a small amount of data to the pre-trained model before the two-stage training. We conduct experiments based on the settings and datasets of the Interspeech2024 Speech Processing Using Discrete Speech Unit Challenge\footnote[1]{\url{https://www.wavlab.org/activities/2024/Interspeech2024-Discrete-Speech-Unit-Challenge}}, which aims to investigate the efficient utilization of SSL model representations to enhance the performance of downstream models employing discrete units as inputs. On the ML-SUPERB dataset \cite{shi23g_interspeech}, the CER relative reduces 44\% compared to the model without any improving methods. Additionally, the performance gap between discrete tokens and continuous representations is significantly reduced. In the best case, the CER for discrete tokens is only 8\% higher than that of continuous representations on the same frontend model.
\vspace{-0.4cm}
\section{Proposed Methods\vspace{-0.35cm}}
\begin{figure}
    \centering
    \includegraphics[width=0.8\textwidth]{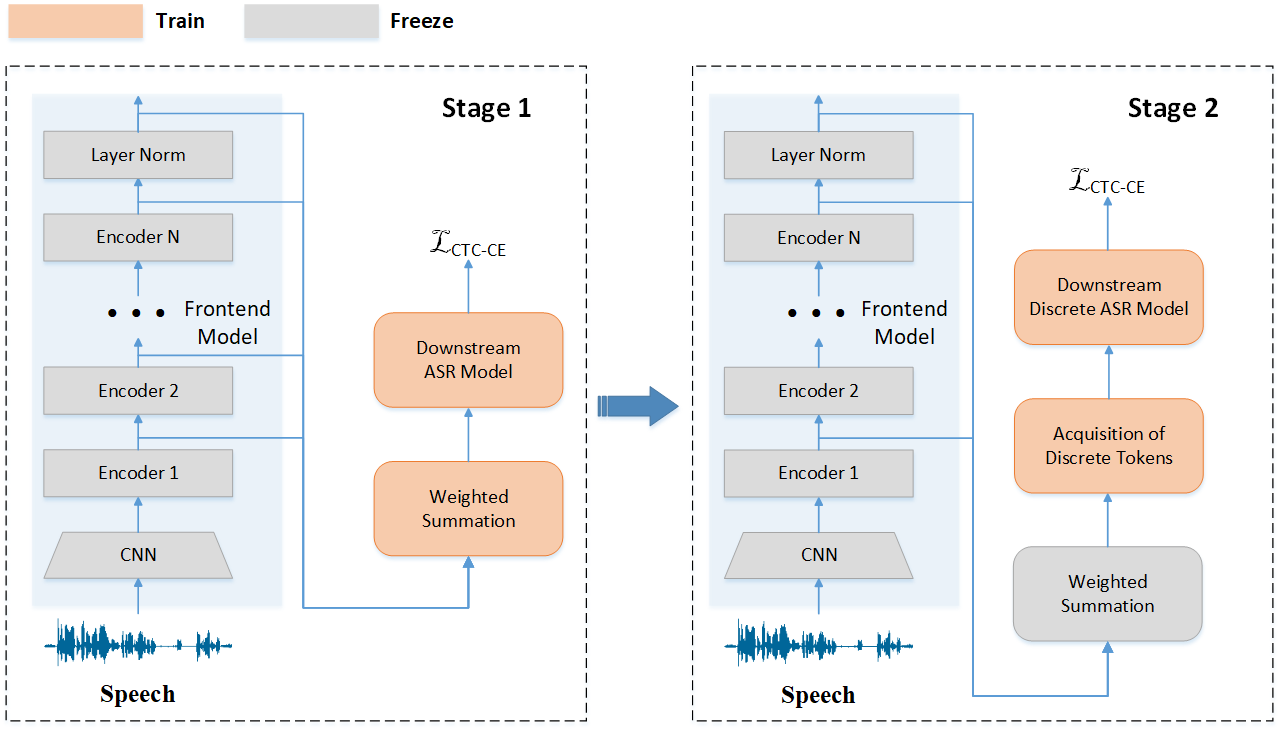}
    \caption{Two-stage discrete speech unit ASR system architecture.} \label{system}
\end{figure}

The architecture illustrated in Figure~\ref{system} comprises three key modules: the frontend model, the acquisition of discrete tokens, and the downstream model. The two-stage training process is structured as follows: the first stage involves the frontend model and downstream tasks, whereas the second stage incorporates all three modules.

\vspace{-0.4cm}
\subsection{Two-stage Training Strategy}
\vspace{-0.3cm}
As mentioned in \cite{shi23g_interspeech}, the representations of different layers contribute differently to different languages, and there are also several layers that have a certain degree of similarity. In order to better utilize the representation differences of each layer and perform well on multilingual data, we choose to extract representations in the form of weighted summation, which is called \emph{weighted\_sum} in this paper.

The weights are learned on a conventional downstream ASR task utilizing continuous representations in the first stage. In order to prevent the model from deviating too much when facing different data distributions or noisy data, we also add the final normalized result as a layer of representation to \emph{weighted\_sum}. Thus, for the pre-trained model, we add the result of the last layer after layer normalization when calculating \emph{weighted\_sum}. For the fine-tuned model, since the last few layers already have word identity, we only select the inter-layer representation. Assume that the model consists of \(l\) encoder blocks, where the representation of the \(i\)-th block is \(h_i\), then the final representation \(h^*\) can be expressed as
\begin{equation}
    h^* = 
    \begin{cases}
        \frac{\sum_{i=1}^{l}e^{\lambda_i}h_i + e^{\lambda_{l+1}}\mathcal{F}(h_{l})}{\sum_{j=1}^{l+1}e^{\lambda_j} } & \text{if}\ \ \ \textit{pre-trained model} \\
        \frac{\sum_{i=1}^{l}e^{\lambda_i}h_i}{\sum_{j=1}^{l}e^{\lambda_j} } & \text{if}\ \ \ \textit{fine-tuned model}
    \end{cases}
    \label{eq1}
\end{equation}
where \(\lambda_i\) are trainable parameters for the \(i\)-th block and \(\mathcal{F}\) denotes the normalization before the final outputs. Usually we use layer normalization, \(\frac{h_{l+1} - \mu }{\sqrt{\sigma^2 -\epsilon  } }\cdot \gamma+\beta\), where \(\mu\) and \(\sigma\) represents the expectation and variance of \(h_{l+1}\), \(\gamma\) and \(\beta\) are offsets.
In the first stage, \(h^*\) participates in the backward propagation. In the second stage, \(\lambda_i\) is frozen and the frontend model only generates different \(h_i\) according to different data, so as to obtain discrete tokens. In the end, we use these discrete tokens to train a discrete ASR model.

Additionally, considering the concentration of the language of pre-trained models intersects with the required language collection but may not overlap, and to make the representation better at encoding word identity, we use a small amount of data to fine-tune the frontend model before the two-stage training process.
\vspace{-0.2cm}
\subsection{Module Description}
\vspace{-0.2cm}
Frontend Model: We mainly use the XLS-R 300M model\footnote[2]{\url{https://huggingface.co/facebook/wav2vec2-xls-r-300m}}, which is called \emph{xlsr} in this paper, to validate our proposed method. \emph{xlsr} is pre-trained on 128 languages and approximately 436K hours of unlabeled speech data, achieving outstanding results on multilingual tasks. It follows Wav2vec2 structure with 24 blocks of encoder and the intermediate hidden dimension is 1024. To demonstrate that our proposed method is also effective on supervised training models, we introduce Whisper large-v3\footnote[3]{\url{https://github.com/openai/whisper}} as another frontend model, referred to as \emph{whisper} in this paper. This model has been trained on 1 million hours of weakly labeled audio and 4 million hours of pseudo-labeled audio collected using Whisper large-v2. \emph{whisper} consists of 32 layers encoder and 32 layers decoder, comprising a total of 1.5 billion parameters.

Acquisition of Discrete Tokens: We train a k-means model using continuous hidden representations with a cluster number of 2000 as reported in \cite{chang2023exploration} and obtain the cluster indices as discrete units using in the ASR model. After obtaining the discrete units through k-means, in order to eliminate redundancy and further improve
computational efficiency, we also apply the method of de-duplication and subword modeling of the discrete units. 
A randomly initialized embedding layer is used before the discrete ASR model.

Downstream Model: We use the joint CTC/attention-based architecture, E\_Branchformer \cite{e_branchformer}, provided by the challenge as the discrete ASR model. The encoder consists of 12 blocks with an intermediate hidden dimension of 1024. The decoder is a 6-layer Transformer with a feed-forward network (FFN) dimension of 2048. We replace the original 2-layer 1D-convolutional block before the encoder layer with a linear layer, and set batch\_bins to 6e7 to match the input length. The CTC weight is set to 0.3, and no language models are used during training or decoding. Training is conducted on one A100 GPU for 100 epochs with gradient accumulation set to 4.

Fine-tuning Choice: Specifically, the \emph{xlsr} model is fine-tuned with the CTC loss using the fairseq toolkit\footnote[4]{\url{https://github.com/facebookresearch/fairseq}}. The peak learning rate is set to 4e-5 and the max update number is 80000. To avoid issues with short multilingual data causing entire sentences to be masked out, we disable masking in both the time and channel dimensions. During the first 10,000 steps, all parameters except the last projection layer are frozen. The best checkpoint during fine-tuning is retained for representation extraction.
\vspace{-0.4cm}
\section{Experiments and Results}
\vspace{-0.3cm}
\subsection{Dataset}
\vspace{-0.3cm}
The challenge provides Librispeech train-clean-100 set (\emph{LS100} in this paper) and ML-SUPERB 1-hour set (\emph{ML-SUPERB} in this paper) for training. In this paper, \emph{LS100} alone is named \emph{train1} and the provided train sets is named \emph{train2}. The dev-\{clean, other\} sets and test-\{clean, other\} sets of Librispeech and test-1h set of ML-SUPERB are used as the test sets. In order to obtain better representation performance, we add the whole Librispeech 960 hour train set (\emph{LS960} in this paper) to conduct fine-tuning on the pre-trained model, which is named \emph{train3}. SpecAugment is used during downstream ASR task. No speed perturbation is applied throughout the experiments. CER is calculated for both Librispeech and ML-SUPERB for consistency.
\vspace{-0.3cm}
\subsection{Fine-tuned Results}
\vspace{-0.3cm}
We fine-tune the \emph{xlsr} model on various datasets and evaluate its performance on test-other and test-1h. During experiment, we find that the default BPE with 6000 subword units fails to completely cover the characters in the training set, resulting in many \(<\)\(unk\)\(>\) tokens and poor performance. We find that increasing the BPE units to 6288 effectively covers all the characters. To further enhance fine-tuning and discrete representations, we also experiment with BPE 6500 and character (char) units. According to Table~\ref{tab:finetune result}, fine-tuning only on \emph{train1} has no improvement on the English dataset, but CER on test-1h increases significantly, indicating a bias towards Librispeeech data. Compared to \emph{train2} provided by the challenge, adding additional Librispeech data can achieve better performance, which is reflected by \emph{xlsr\_train3} outperforming \emph{xlsr\_train2}. The fully covered BPE 6500 setting performs significantly better than the default, whereas the result of \emph{xlsr\_train3\_char} is slightly better than that of \emph{xlsr\_train3\_bpe6500}. This may be due to the relatively small proportion of alphabet-based languages, such as English and French, in the test sets. \emph{whisper\_train3} is fine-tuned using BPE 6500 because this setting yields the best results for \emph{whisper} model. We add a projection layer after the encoder blocks of \emph{whisper} and using the CTC loss instead of the cross entropy (CE) loss with decoder blocks. The results of \emph{whisper} are slightly worse than the results of \emph{xlsr}.
\vspace{-0.5cm}
\begin{table*}
    \caption{CER (\%) of fine-tuned models using different corpus.}
    \label{tab:finetune result}
    \centering
        \begin{tabular}{cccc}
        \hline
        \textbf{Model} & \textbf{Corpus} & \textbf{Test-other} & \textbf{Test-1h} \\
        \hline
        xlsr\_train1         & LS100         & 7.5 & 70.9 \\
        xlsr\_train2 & ML-SUPERB + LS100 & 7.4 & 27.8 \\
        xlsr\_train3 & ML-SUPERB + LS960 & 4.3 & 23.6 \\
        xlsr\_train3\_bpe6288 & ML-SUPERB + LS960 & 3.2 & 16.3 \\
        xlsr\_train3\_bpe6500 & ML-SUPERB + LS960 & 3.5 & 18.4 \\
        xlsr\_train3\_char & ML-SUPERB + LS960  & 3.3 & 16.4 \\
        whisper\_train3          & ML-SUPERB + LS960 & 6.5 & 23.5 \\
        \hline
        \end{tabular}
\end{table*}
\vspace{-0.9cm}
\subsection{Continuous Representation Results}
\vspace{-0.2cm}
During the stage of extracting continuously representation for ASR task, we only use \emph{train\_2} so as to be consistent with the challenge baseline. Parameters of the pre-trained or fine-tuned model used as the frontend model are frozen, and only the weights of the \emph{weighted\_sum} are updated. All encoder layers participate in the \emph{weighted\_sum} calculation, and \emph{xlsr} pre-trained model and \emph{whisper} also add additional normalized layer results. The results which serve as a top line for discrete tokens are shown in Table~\ref{tab:continuous result}. Compared with the direct usage of \emph{xlsr} representation, the CER after fine-tuning reduces significantly. Specifically, \emph{xlsr\_train3\_char} has a relative decrease of 45\% on test-other set and a relative decrease of 12\% on test-1h set. However, the fine-tuned Whisper model \emph{whisper\_train3} shows consistent performance with \emph{whisper} on the Librispeech dataset, with only a significant decrease on CER observed on the test-1h set. Considering the fine-tuned result in Table~\ref{tab:finetune result}, it also shows that the good performance of Whisper on multilingual ASR task depends more on the decoder blocks.

\vspace{-0.5cm}
\begin{table*}
    \caption{CER (\%) of downstream ASR task using continuous representations.}
    \label{tab:continuous result}
    \centering
        \begin{tabular}{cccccc}
        \hline
        \textbf{Frontend Model} & \textbf{Dev-clean} & \textbf{Dev-other} & \textbf{Test-clean} & \textbf{Test-other} & \textbf{Test-1h} \\
        \hline
        whisper              & 3.0 & 4.2 & 1.9 & 4.3 & 22.3 \\
        xlsr           & 2.9 & 7.6 & 2.7 & 7.5 & 17.7 \\
        xlsr\_train3 & 1.5 & 4.1 & 1.3 & 4.1 & 16.1 \\
        xlsr\_train3\_bpe6288 & 1.1 & 3.2 & 1.1 & 3.1 & 15.9 \\
        xlsr\_train3\_char & 1.2 & 3.2 & 1.1 & 3.2 & 15.6 \\
        whisper\_train3  & 2.8 & 4.5 & 2.3 & 4.6 & 14.3 \\
        \hline
        \end{tabular}
\end{table*}

\vspace{-1.0cm}
\subsection{Discrete ASR Task}
\begin{figure}
    \includegraphics[width=\textwidth]{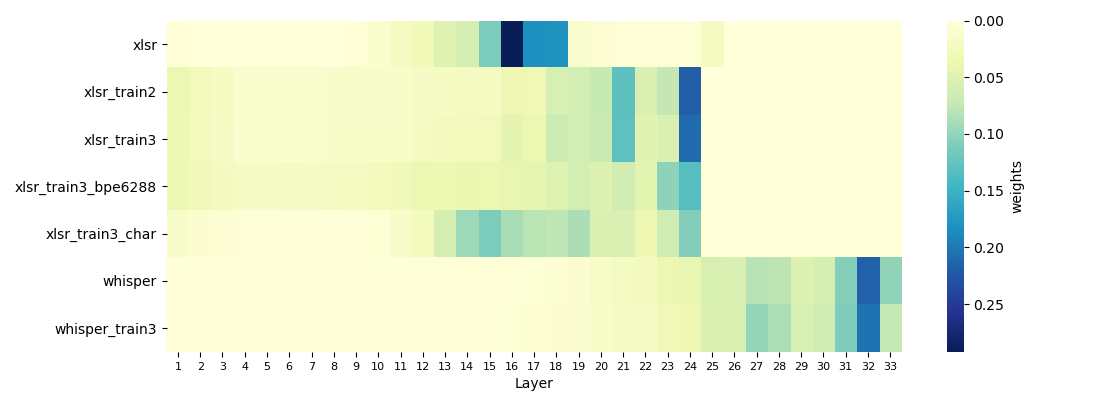}
    \caption{Weight distribution of different models on \emph{train\_2}. The XLSR pre-trained model has 25 layers of weights, whereas its fine-tuned version has 24 layers. To ensure alignment in the plot with the weights of the Whisper model (33 layers), we set the weights of the remaining layers in the XLSR-based models to 0.} \label{heatmap}
\end{figure}

\vspace{-0.2cm}
Similarly, we only train discrete ASR models on \emph{train\_2} utilizing discrete tokens. To better understand the contribution of representation at each layer across different models, we visualize the weights of these models as shown in Figure~\ref{heatmap}. For XLSR pre-trained model \emph{xlsr}, the weights tend to be concentrated in the middle layers, while for XLSR-based fine-tuned models and Whisper models, the weights are concentrated in the last few layers. This confirms that after fine-tuning, the performance of the representation layer in XLSR-based models aligns with that of supervised training models like Whisper, with language information predominantly concentrated in the final layers.

For discrete token experiments on individual layers, we extracted the representations from the layer with the highest weight: the 17-th layer of \emph{xlsr} and the 24-th layer of \emph{xlsr\_train3}. The usage of \emph{weighted\_sum} is incorporated from the continuous representation using Eq~\ref{eq1}. We reproduce the baseline results on the WavLM\footnote[5]{\url{https://huggingface.co/microsoft/wavlm-large}} pre-trained model given in the challenge. As shown in Table~\ref{tab:discrete_result}, utilizing the representation obtained through \emph{weighted\_sum} on \emph{xlsr} outperforms both the Librispeech and ML-SUPERB test sets compared to solely specifying the 17-th layer representation, although still falling short of the the WavLM baseline. Regardless of whether using a single-layer representation or a \emph{weighted\_sum} representation, there is a relative improvement of about 50\% on the Librispeech test sets and a relative improvement of about 20\% on ML-SUPERB test-1h set compared to directly using the representations of the pre-trained \emph{xlsr} model. Additionally, comparing the discrete token results with the continuous representation results on XLSR-based models, the proposed method significantly reduces the gap between these two results, with even a relative difference of only 8\% on \emph{xlsr\_train3\_char}.


\vspace{-0.5cm}
\begin{table*}
    \caption{CER (\%) of downstream discrete ASR task using discrete tokens.}
    \label{tab:discrete_result}
    \centering
    \scalebox{0.85}{
        \begin{tabular}{ccccccc}
        \hline
        \textbf{Frontend Model} & \textbf{Layer Usage} &\textbf{Dev-clean} & \textbf{Dev-other} & \textbf{Test-clean} & \textbf{Test-other} & \textbf{Test-1h} \\
        \hline
        wavlm  & 21            & 1.6 & 3.4 & 1.5 & 3.3 & 22.9 \\
        xlsr   & 17            & 3.6 & 9.7 & 3.4 & 10.1 & 30.7 \\
        xlsr   & weighted\_sum & 3.4 & 9.3 & 3.4 & 9.5 & 25.3 \\
        whisper      & weighted\_sum & 5.1 & 8.1 & 5.0 & 8.2 & 24.0 \\
        xlsr\_train3 & 24    & 1.9 & 4.6 & 1.8 & 4.5 & 22.3 \\
        xlsr\_train3 & weighted\_sum & 1.7 & 4.2 & 1.6 & 4.1 & 21.2 \\
        xlsr\_train3\_bpe6288 & 24 & 1.2 & 3.4 & 1.2 & 3.3 & 18.0 \\
        xlsr\_train3\_bpe6288 & weighted\_sum & 1.2 & \textbf{3.4} & 1.2 & 3.3 & 17.4 \\
        xlsr\_train3\_char & weighted\_sum & \textbf{1.2} & 3.5 & \textbf{1.2} & \textbf{3.3} & \textbf{16.9} \\
        whisper\_train3 & weighted\_sum & 4.2 & 7.2 & 4.1 & 7.4 & 23.6 \\
        \hline
        \end{tabular}
    }
\end{table*}
\vspace{-0.7cm}

\vspace{-0.3cm}
\section{Conclusion}
\vspace{-0.3cm}
In this paper, we propose a two-stage training strategy along with a fine-tuning stage to improve the performance of discrete tokens for multilingual ASR tasks. 
Experiments demonstrates that this approach significantly reduces the performance gap between discrete tokens and continuous representations, which achieves the first place among all the single-system results on the leaderboard of the challenge. In future work, we will explore methods to obtain discrete tokens that retain all relevant information for the downstream task, maximizing the benefits of training with discrete tokens.

%
%
%
\bibliographystyle{splncs04_unsort}
\bibliography{mybib}
%






\end{document}